\documentclass{iaus}
\usepackage{graphicx}

  \checkfont{eurm10}
  \iffontfound
    \IfFileExists{upmath.sty}
      {\typeout{^^JFound AMS Euler Roman fonts on the system,
                   using the 'upmath' package.^^J}%
       \usepackage{upmath}}
      {\typeout{^^JFound AMS Euler Roman fonts on the system, but you
                   dont seem to have the}%
       \typeout{'upmath' package installed. iaus.cls can take advantage
                 of these fonts,^^Jif you use 'upmath' package.^^J}%
      }
  \else
  \fi

  \checkfont{msam10}
  \iffontfound
    \IfFileExists{amssymb.sty}
      {\typeout{^^JFound AMS Symbol fonts on the system, using the
                'amssymb' package.^^J}%
       \usepackage{amssymb}%

      }{}
  \fi

  \IfFileExists{amsbsy.sty}
    {\typeout{^^JFound the 'amsbsy' package on the system, using it.^^J}%
     \usepackage{amsbsy}}
    {}

\newsavebox{\astrutbox}
\sbox{\astrutbox}{\rule[-5pt]{0pt}{20pt}}

\def\Msun{M$_{\odot}\,$}
\def\solm{M$_{\odot}\,$}

\title[The Interplay among Black Holes, Stars and ISM in Galactic 
       Nuclei]{Nuclear spirals: gas in asymmetric galactic
       potential with a massive black hole}

\author[Witold Maciejewski]%
{Witold Maciejewski}

\affiliation{Obserwatorium Astronomiczne Uniwersytetu Jagiello\'{n}skiego,
Orla 171, 30-244 Krak\'{o}w, Poland}

\pubyear{2004}
\volume{222}
\pagerange{1--8}
\date{?? and in revised form ??}
\jname{The Interplay among Black Holes, Stars and ISM \\in Galactic Nuclei}
\editors{Th. Storchi Bergmann, L.C. Ho \& H.R. Schmitt, eds.}
\begin{document}

\maketitle

\begin{abstract}
Nuclear spirals can provide a wealth of information about the nuclear 
potential in disc galaxies. They form naturally as a gas response to
non-axisymmetry in the gravitational potential, even if the degree of this
asymmetry is very small. Linear wave theory well describes weak nuclear 
spirals, but stronger asymmetries in the potential induce waves beyond 
the linear regime, which appear as spiral shocks. If a central massive 
black hole (MBH) is present, spiral shocks can extend all the way to its 
immediate vicinity, and generate gas inflow up to 0.03 \Msun/yr. This 
coincides with the accretion rates needed to power local Active Galactic 
Nuclei.
\end{abstract}

\firstsection 
\section{Introduction}
Recent high-resolution maps of galactic centres (Regan \& Mulchaey 1999, 
Martini \& Pogge 1999, Pogge \& Martini 2002, Martini et al. 2003 a,b) 
indicate that 50\% to 80\% of disc galaxies posses {\it nuclear spirals}
in their innermost hundreds of parsecs. It has been 
proposed that nuclear spirals may be related to the fueling of Seyfert 
activity (Regan \& Mulchaey 1999). Martini \& Pogge (1999) showed that 
nuclear spirals are not self-gravitating, and that they are likely to be 
shocks in nuclear gas discs.

Here I investigate the dynamics of nuclear spirals which are density 
waves generated in gas by a rotating potential. 
In particular, I am interested in how the gas flow in the nucleus is modified 
by the presence of a central massive black hole (MBH) or a density cusp. 
I construct hydrodynamical models of nuclear spirals with the code that
follows the gas flow on a polar grid in a fixed asymmetric potential of 
a weak oval or of a strong bar. Potentials with a $10^8$ \solm\ massive 
black hole (MBH) or without a MBH (e.g. constant-density core) are 
considered here.
Self-gravity in gas is not taken into account, and I assume isothermal 
equation of state for 'hot gas' (sound speed $c_S = 20$ km/s). In order to
minimize spurious inflow resulting from the inner boundary on the polar grid, 
I impose the reflection boundary condition there. Full account of this work 
will be published elsewhere (Maciejewski 2004 a,b).

\begin{figure}
\vspace{5mm}
\centering
\resizebox{13cm}{!}{\includegraphics{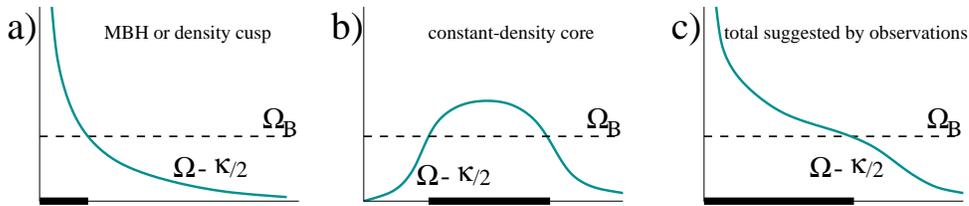} }
\caption{Typical frequency curves (frequency as a function of radius) for 
gravitational potentials of: {\bf a)} a MBH or a central density cusp,
{\bf b)} continuous mass distribution with a constant-density core, {\bf c)} 
a typical composite supported by common appearance of nuclear spirals. Regions
where nuclear spirals can propagate are indicated by thick lines on the
horizontal axis.}
\end{figure}

\section{Zones of propagation for nuclear spirals}
Nuclear spirals can be treated as waves in gas (Goldreich \& Tremaine 1979,
Englmaier \& Shlosman 2000, Maciejewski 2004a), and in the tight-winding 
limit they follow the dispersion relation for wavenumber $k$,
$(\Omega + \kappa/m - \Omega_B) (\Omega - \kappa/m - \Omega_B) = 
k^2 c_S^2 / m^2$,
when excited by a tumbling $m$-fold pattern of frequency $\Omega_B$=const.
In the formula above, $\Omega$ is the angular velocity in the disk, and
$\kappa$ the epicyclic frequency. Note that for a twofold pattern (e.g. 
a bar) in the galaxy centre $\Omega + \kappa/2 - \Omega_B$ is always 
positive, therefore $\Omega - \kappa/2 - \Omega_B$ must be positive in
order for the spiral wave to propagate. Therefore the presence of nuclear 
spirals in galaxies can tell us about the potential in the galaxy centre:
since $\Omega_B$=const, then if the nuclear spiral extends to the galaxy 
centre, $\Omega - \kappa/2$ must be large there, which is characteristic
for a MBH or a central density cusp (Fig.1). To the contrary, constant-density
core (linearly rising inner rotation curve) implies 
$\Omega - \kappa/2 \equiv 0$, and therefore prohibits propagation of
any nuclear spiral wave.

\section{Nuclear spirals in a weak oval}
In order to check how weak departures from axisymmetry in the stellar
gravitational potential are sufficient to create nuclear spirals in gas, 
I built models for the potential, whose departure from axisymmetry
is 10 times smaller than that for a strong bar studied by Maciejewski et al.
(2002). This potential contains a weak oval distortion rather than a bar, 
with the maximum ratio of tangential to radial force of 2\%. As can be
seen in Fig.2, such a weak asymmetry is sufficient to generate a nuclear 
spiral with the arm/inter-arm density contrast $\sim 2$. However, velocity 
perturbations in the spiral are small, therefore it is a wave, not a shock,
in gas. As explained by the linear theory, the nuclear spiral extends
inwards from the outer inner Lindblad resonance (ILR) to the inner ILR,
or to the galaxy centre in the absence of this last one (Fig.2, middle 
columns). Note that despite the assumed relatively high 
velocity dispersion in gas ($c_S = 20$ km/s), to which the pitch angle of
the spiral is proportional, the nuclear spiral in Fig.2 appears tightly 
wound.

\begin{figure}
\vspace{-8mm}
\centering
\resizebox{14cm}{!}{\includegraphics{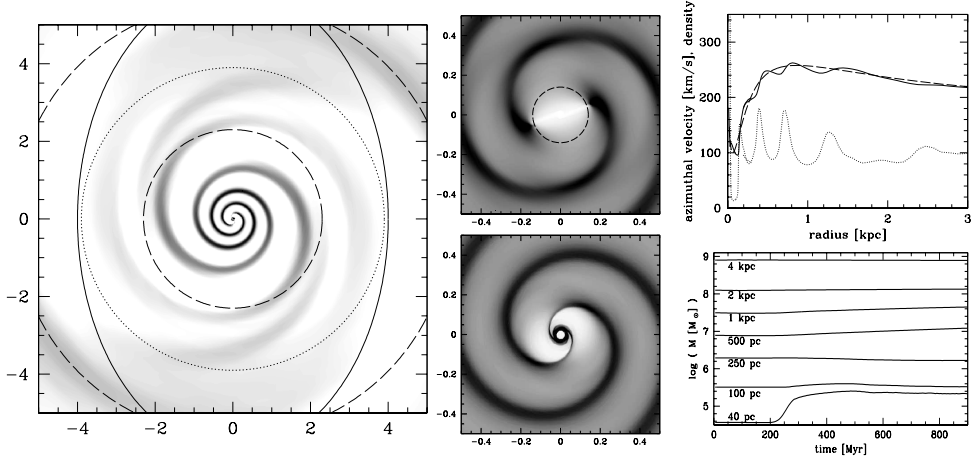} }
\vspace{-125mm}
\caption{{\bf Left:} Snapshot of gas density in a model with a weak oval 
asymmetry in the gravitational potential once the
morphology of the flow has stabilized. Darker color indicates larger 
densities. The solid ellipse outlines the oval, and the dashed circles 
mark the outer ILR at 2.3 kpc and the corotation at 5.6 kpc.
{\bf Top-central:} Zoom into the central kiloparsec of the model with a 
constant-density core and without a MBH. The dashed circle marks the inner
ILR.
{\bf Bottom-central:} Zoom into the central kiloparsec of the model with a
MBH. The inner grid boundary is at the radius of 20 pc. 
{\bf Top-right:} Azimuthal velocity (solid line) plotted against the 
unperturbed rotation curve (dashed line) as a function of radius, and the 
radial density profile (dotted line).
{\bf Bottom-right:} Mass accumulated within various radii as a function 
of time.}
\end{figure}

Although the nuclear spiral generated by a weak oval is a strong perturbation 
in gas density, no inflow is triggered at any radial scale (Fig.2, 
bottom-right). Only at
the innermost radii an episodic drop of $\sim 10^5$ \solm occurs during 
formation of the spiral. This corresponds to the inflow of $\sim 10^{-3}$ 
\solm/yr, and may be only relevant to the feeding of weakest AGN. However,
it may gain importance if nuclear spirals re-appear as a response to drivers
like passing Globular Clusters or Giant Molecular Clouds.

\section{Nuclear spirals in a strong bar}
Nuclear spiral in a strong bar (here I use the potential from Maciejewski et 
al. 2002) propagates inwards from the inner ends of straight shocks in the 
bar (Fig.3). Contrary to the spiral in a weak oval, it appears as
a slight density enhancement, but with large negative $div^2${\bf v},
hence it is {\it a spiral shock}. Its strength equals that of the principal 
straight shock in the bar, and the departures from circular velocity in the 
spiral are large (Fig.3, middle-top). Pitch angle of the spiral shock
is consistently larger than what the linear theory predicts (Fig.3, 
middle-bottom). Its morphology does not change with time, but the
density in the spiral increases, since the gas transported inwards
by the straight principal shock in the bar accumulates in the nuclear spiral.

This accumulation of gas in the nuclear spiral is responsible for secular
inflow in the spiral shock. Its timescale is much longer than that in the
straight principal shock in the bar, but with time its amplitude reaches 
that in the straight shock: 0.7 \solm/yr (Fig.3, right panels). At the 
radius of 500 pc it 
happens $\sim 0.4$ Gyr after formation of the bar (2.5 rotations of the bar).
The same large inflow will likely occur at smaller radii, but after still
longer times, and evolutionary changes in the galaxy may override its 
importance. Nevertheless, in addition to this secular inflow, another form 
of inflow takes place around a MBH if it is present in the galaxy centre. 
This inflow reaches 0.03 \solm/yr after 0.5 Gyr, comparable to what is needed
to sustain activity of local Seyfert galaxies (Peterson et al. 1997).

\begin{figure}
\vspace{-8mm}
\centering
\resizebox{14cm}{!}{\includegraphics{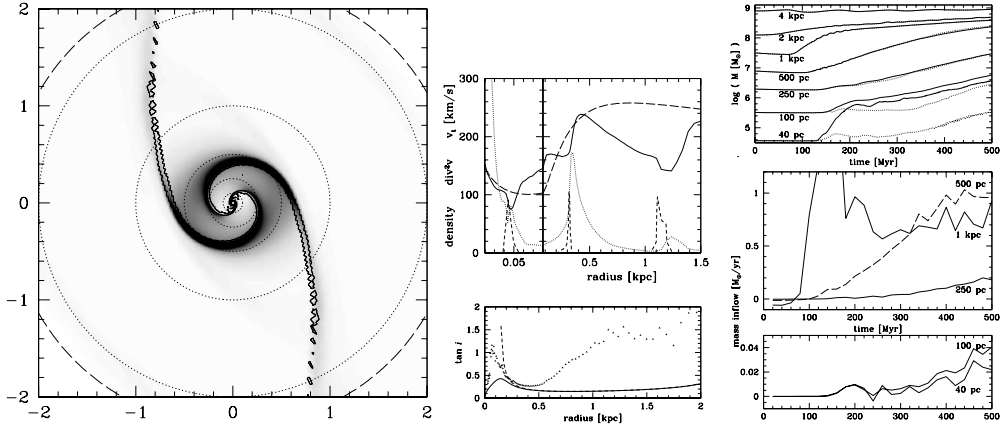} }
\vspace{-125mm}
\caption{{\bf Left:} Snapshot of gas density (greyscale), and of 
$div^2 {\bf v}$ (for $div {\bf v} < 0$, contours) in a model 
with a MBH and a strong bar, once the morphology of 
the flow has stabilized. Overplotted are dotted circles at radii where 
the inflow has been measured.
{\bf Top-central:} Azimuthal velocity (solid line) with the rotation curve 
(long-dashed line), radial density profile (dotted line), and 
$div^2 {\bf v}$ (short-dashed line). The plot has two adjacent parts drawn 
in different radial scales: from 20 pc to 80 pc, and from 80 pc to 1.5 kpc.
{\bf Bottom-central:} Tangens of the pitch angle of the shock. Filled 
triangles mark the values measured in the hydrodynamical model, while lines
are the linear prediction for a potential with a central MBH (solid),
and without it (dashed).
{\bf Top-right:} Mass accumulated within various radii as a function of 
time for model with a MBH (solid line) and without it (dotted line).
{\bf Bottom-right:} Mass inflow as a function 
of time in model with a MBH. Note the small, but not negligible inflow
triggered at the innermost radii after the arrival of the spiral shock there
(at about 130 Myr).
}
\end{figure}

\section{Confronting observations}
Models presented here can be related to
the morphological classification of the nuclei, and the statistics of nuclear
activity in a recent sample of active and inactive galaxies (Martini et al. 
2003 a,b). In the data, nuclear spirals classified as 'grand design' and 
'loosely wound' occur in 60\% of active galaxies and in 20\% 
of inactive ones. 'Grand design' class directly corresponds to the spiral shock
in the strong bar modeled here, and is not observed in unbarred galaxies.
'Loosely wound' class indicates large pitch angle of the spirals, which models
associate with the spiral shock. The observational fact that both these 
classes occur
more often in active than in non-active galaxies supports the idea that the
spiral shock causes gas inflow large enough to feed the central MBH.
Nuclear spirals in the 'tightly wound' class of Martini 
et al. tend to avoid galaxies classified as barred. In the 
models, the nuclear spiral in a weak oval can follow the linear theory 
for longer than in a strong bar, and therefore it can appear as tightly wound
there.

\section{Conclusions}
Nuclear spirals are easily generated even by small asymmetries in the galactic
potential, but they cannot form in constant-density cores. Some nuclear 
spirals can feed the central MBH, but observations of gas kinematics are 
essential in testing it.

\end{document}